\documentclass[a4paper,fleqn,usenatbib]{mnras}

\usepackage{newtxtext,newtxmath}

\usepackage[T1]{fontenc}
\usepackage{ae,aecompl}

\usepackage{graphicx}	
\usepackage{amsmath}	
\usepackage{amssymb}	
\usepackage{tikz}

\newcommand{\hi}{H\,{\sc i}}
\newcommand{\htwo}{\mbox{H}_2}

\newcommand{\kms}{\mbox{$\rm km\, s^{-1}$}}

\usetikzlibrary{arrows.meta}
\tikzset{%
  >={Latex[width=2mm,length=2mm]},
            base/.style = {rectangle, rounded corners, draw=black,
                           minimum width=4cm, minimum height=1cm,
                           text centered, font=\sffamily},
  loadData/.style = {base, fill=blue!30},
       startstop/.style = {base, fill=red!30},
    SourceFinding/.style = {base, fill=green!30},
         DataReduction/.style = {base, minimum width=2.5cm, fill=orange!15,
                           font=\ttfamily},
}

\title[Molecular Column Density Function]{ALMACAL VI: Molecular gas mass density across cosmic time via a blind search for intervening molecular absorbers}

\author[A. Klitsch et al.]{
Anne~Klitsch$^{1, 2}$,\thanks{E-mail: aklitsch@eso.org}
C\'eline~P\'eroux$^{1, 3}$,
Martin~A.~Zwaan$^{1}$,
Ian~Smail $^{2}$,
Dylan~Nelson$^{4}$,
\newauthor{
Gerg\"o~Popping$^{5}$,
Chian-Chou~Chen$^{1}$,
Benedikt~Diemer$^{6}$,
R.~J.~Ivison$^{1, 7}$,}
\newauthor{
James~R.~Allison$^{8}$,
S\'ebastien~Muller$^{9}$,
A.~Mark~Swinbank $^{2}$,
Aleksandra~Hamanowicz$^{1}$,}
\newauthor{
Andrew~D.~Biggs$^{1}$,
Rajeshwari~Dutta$^{1}$}
\\
$^{1}$European Southern Observatory, Karl-Schwarzschild-Str. 2, 85748 Garching near Munich, Germany\\
$^{2}$Centre for Extragalactic Astronomy, Durham University, Department of Physics, South Road, Durham DH1 3LE, UK\\
$^{3}$Aix Marseille Univ, CNRS, LAM, (Laboratoire d'Astrophysique de Marseille), UMR 7326, 13388, Marseille, France\\
$^{4}$Max-Planck-Institut f\"ur Astrophysik, Karl-Schwarzschild-Str. 1, 85741 Garching, Germany\\
$^{5}$Max-Planck-Institut f\"ur Astronomie, K\"onigstuhl 17, 69117 Heidelberg, Germany\\
$^{6}$Institute for Theory and Computation, Harvard-Smithsonian Center for Astrophysics, 60 Garden St., Cambridge, MA 02138, USA\\
$^7$ Institute for Astronomy, University of Edinburgh, Royal Observatory, Blackford Hill, Edinburgh EH9 3HJ, UK\\
$^{8}$Sub-Dept. of Astrophysics, Department of Physics, University of Oxford, Denys Wilkinson Building, Keble Rd., Oxford, OX1 3RH, UK\\
$^{9}$Department of Space, Earth and Environment, Chalmers University of Technology, Onsala Space Observatory, 43992 Onsala, Sweden\\
}

\date{Accepted XXX. Received YYY; in original form ZZZ}

\pubyear{2019}

\begin{document}
\label{firstpage}
\pagerange{\pageref{firstpage}--\pageref{lastpage}}
\maketitle

\begin{abstract}
We are just starting to understand the physical processes driving the dramatic change in cosmic star-formation rate between $z\sim 2$ and the present day. A quantity directly linked to star formation is the molecular gas density, which should be measured through independent methods to explore variations due to cosmic variance and systematic uncertainties. We use intervening CO absorption lines in the spectra of mm-bright background sources to provide a census of the molecular gas mass density of the Universe. The data used in this work are taken from ALMACAL, a wide and deep survey utilizing the ALMA calibrator archive. 
While we report multiple Galactic absorption lines and one intrinsic absorber, no extragalactic intervening molecular absorbers are detected. However, thanks to the large redshift path surveyed ($\Delta z=182$), we provide constraints on the molecular column density distribution function beyond $z\sim 0$. In addition, we probe column densities of N(H$_2$) > 10$^{16}$ atoms~cm$^{-2}$, five orders of magnitude lower than in previous studies. We use the cosmological hydrodynamical simulation IllustrisTNG to show that our upper limits of $\rho ({\rm H}_2)\lesssim 10^{8.3} \text{M}_{\sun} \text{Mpc}^{-3}$ at $0 < z \leq 1.7$ already provide new constraints on current theoretical predictions of the cold molecular phase of the gas.
These results are in agreement with recent CO emission-line surveys and are complementary to those studies. The combined constraints indicate that the present decrease of the cosmic star-formation rate history is consistent with an increasing depletion of molecular gas in galaxies compared to $z\sim 2$. 
\end{abstract}




\section{Introduction}

Understanding the efficiency of converting baryons into stars is a challenge in studies of galaxy formation and evolution. 
The star-formation rate history (SFH) is well established from observations of star-forming galaxies across cosmic time at infrared, ultra violet, submillimetre, and radio wavelengths \citep[][and references therein]{Madau2014}. The star-formation rate (SFR) density increased at high redshift, reached a peak at around $z \sim 2$ and decreased until today \citep[see][for a review]{Madau2014}. Which physical processes are driving this dramatic change and their relative importance represent two of the main unanswered questions of modern astrophysics. Whether this is due to a lack of cold gas supply, or a lower efficiency in converting the gas into stars, or to the presence of strong outflows preventing the infall of new cold material, is still debated \newline \citep[e.g.][and references therein]{Madau2014, Katsianis2017}. Simple expectations would have the SFH mirror the cold gas evolution, as gas is being consumed by star formation \citep[e.g.][]{Putman2017, Driver2018}. The atomic gas density, $\Omega_{\rm HI}$, is the original reservoir of gas for star formation and is indeed well constrained locally and at $z > 2$. Most recent results, however, indicate a mild evolution in $\Omega_{\rm HI}$ with cosmic lookback time \citep[e.g.][]{Noterdaeme2009, Zafar2013, Rhee2018}. Neutral hydrogen provides the essential reservoir, but it has to cool and transform to the molecular phase in order to provide the necessary conditions for star formation. Further studies using damped Lyman~$\alpha$ absorbers as well as \hi~21~cm absorption traced with the Square Kilometre Array (SKA) path finder observations provide important clues on physical state of the atomic gas and the neutral  inter-stellar medium (ISM) physics \citep{Kanekar2014a, Allison2016}.  However, a direct probe of the fuel for star formation has to come from measurements of the molecular phase of the gas. 

 In order to probe this essential phase of baryons over cosmic time, a number of deep cosmological surveys for CO emission lines have been conducted. The first study used the  Institute de Radioastronomie Millim\'etrique (IRAM) Plateau de Bure Interferometer to perform molecular line scans in the Hubble Deep Field North and provided upper limits on the cosmic molecular gas mass density $\Omega(\text{CO})$ \citep{Decarli2014, Walter2014}. More recently, the ALMA Spectroscopic Survey in the Hubble Ultra Deep Field (ASPECS), provided the first measurements of $\Omega(\text{CO})$ at redshift $0 < z < 4.5$ \citep{Decarli2016, Decarli2019}. The CO Luminosity Density at High Redshift (COLDz) survey \citep{Riechers2018} undertaken with the VLA offers first indications of the molecular mass density at high redshift ($z\sim$~2--3 and $z \sim$~5--7). An alternative approach using the dust mass as a tracer of the molecular gas mass is presented by \citet{Scoville2017}. The fact that multiple transitions of CO at different redshifts can be searched in a given observed frequency setting greatly increases the redshift path, and hence the searched volume. These emission-line surveys are especially sensitive to the high-mass end of the molecular gas mass function. However, such observations require large investments in telescope time, and since typically only a small contiguous area is covered, the results are prone to cosmic variance effects.
 
Four intervening molecular absorbers have been detected in targeted surveys of strongly lensed systems and galaxy merger pairs that were known to show \hi\ absorption \citep[e.g.][]{Wiklind1995, Kanekar2005, Wiklind2018, Combes2019}. Only the molecular absorber towards PKS 1830-211 was detected before any other lines were known \citep{Wiklind1996a}. In addition, associated molecular absorption lines have been found in three intermediate-redshift AGN \citep{Wiklind1994, Wiklind1996, Allison2018a}. Similar, intrinsic absorption is observed more frequently in low redshift AGNs \citep[e.g.][]{Tremblay2018a, Maccagni2018, Rose2019} and in high-redshift submillimetre galaxies (SMGs) \citep[e.g.][]{George2014, Falgarone2017, Indriolo2018}. 
\citet{deUgartePostigo2018} also reported CO absorption lines against Gamma Ray Bursts (GRBs) observed with Atacama Large Millimetre and Submillimetre Array (ALMA). 
However, to measure the cosmic molecular gas mass density in an unbiased way, blind detections of intervening molecular absorbers are required.

Here, we present a complementary approach to probing the molecular phase of the gas and its evolution with cosmic time free from cosmic variance issues. In analogy with studies at optical wavelengths \citep[e.g.][]{Peroux2003, Zafar2013}, we use (sub)mm-bright ($\sim 0.1 - 3$~Jy) background sources to probe intervening molecular absorption lines.  Moving from optical to mm wavelengths has the advantage that this study is not affected by dust attenuation in the quasar spectra which might be expected for molecular absorbers. Therefore, by choosing (sub)mm-bright background objects we are not biased against potentially dusty absorbers. Furthermore, tracing molecular absorption offers a measurement of the cosmic molecular gas density free from cosmic variance.

A similar ``blind'' study was performed at lower frequencies using the Green Bank Telescope \citep{Kanekar2014}. The authors surveyed a redshift path, defined as the sum of the redshift intervals covered by the individual spectra ($\Delta z = \sum_i (z_{\rm max} - z_{\rm min})$), of $\Delta z\sim 24$ at $0.81 < z < 1.91$ and did not detect any molecular absorbers with N(H$_2$) $\geq 3 \times 10^{21}$. 
In the present work, we perform a ``blind'' search for CO absorbers against a large sample of (sub)mm bright background galaxies. These objects are all 880 ALMACAL targets observed up until December 2018. ALMACAL consists of observations of a large sample of bright, compact sources \citep[generally blazars, see][]{Bonato2018} which are used as calibrator sources for ALMA. 
These calibrators are ideal targets for an unbiased search for CO absorbers for two main reasons. First, the total integration time spent on ALMACAL sources is $>1500$ hours, orders of magnitude more than what would be attainable in a targeted ALMA survey programme for intervening absorption lines. Secondly, since the calibrators are distributed all over the sky observable with ALMA, it is possible to quantify the effect of cosmic variance. 
In addition, the sensitivity of the absorption survey is independent of redshift and solely relies on the brightness of the unrelated background sources. Using absorption lines we are able to reach low gas column densities, providing us with a more complete and unbiased (with respect to excitation conditions) view of the molecular gas content of the Universe over cosmic time.

There are also several caveats using this blind absorption line approach. For example, a single identification of an absorption line cannot uniquely linked to a species, and hence not to the column density and the redshift of the absorber. However, since CO is a much stronger absorber than all other molecular species we can safely assume that any detected absorption line is tracing CO. Furthermore, unlike emission-line surveys in well-studied cosmological deep fields, our survey does not have the luxury of extensive ancillary data that can be used to identify source redshifts. Our sample of quasars is magnitude limited and therefore susceptible to effects of gravitational magnification. First, the probability of finding quasars with absorbers is increased by the flux boosting from gravitational lensing by the absorber. Second, the solid angle behind absorbers is gravitationally enlarged diluting the flux of the background quasar. \citet{Menard2003} find indeed an excess of bright quasars with absorbers. Furthermore, ALMA calibrators are selected to be (sub)mm bright and have therefore redshifts of $z \leq 3$.

The paper is organised as follows: the ALMACAL survey including the optimised data reduction for this data-intensive project is presented in Sec.~\ref{SecDataRed}, in Sec.~\ref{SecBlindDetec} we describe the absorption line search as well as the derivation of the  limits on the CO column density distribution function from our observations and the molecular gas column density distribution function from the IllustrisTNG simulation, in Sec.~\ref{SecDiscussion} we present our limits on the molecular gas mass density evolution as a function of cosmic time and in Sec.~\ref{SecConclusions} we summarize our conclusions. Throughout this paper we adopt a $\Lambda$CDM cosmological model with $H_0 = 70 {\rm km s^{-1} Mpc^{-1}}$, $\Omega_{\rm m} = 0.3$ and $\Omega_{\Lambda} = 0.7$.

\section{ALMACAL observations and data reduction}
\label{SecDataRed}

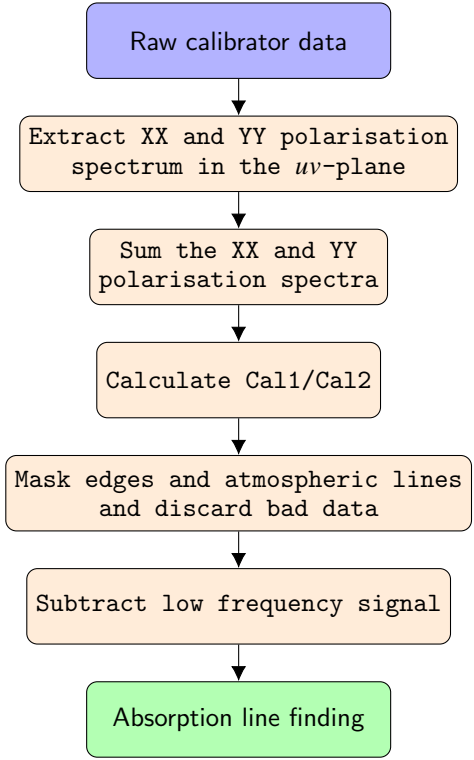
\begin{figure}
\centering
\begin{tikzpicture}[node distance=1.5cm,
    every node/.style={fill=white, font=\sffamily}, align=center]
    \large
  \node (load)             [loadData]              {Raw calibrator data};
  \node (extractSpectrum)      [DataReduction, below of=load]   {Extract XX and YY polarisation \\ spectrum in the $uv$-plane};
  \node (sumSpectrum)     [DataReduction, below of=extractSpectrum]   {Sum the XX and YY \\ polarisation spectra};
  \node (divideSpectra)     [DataReduction, below of=sumSpectrum]   {Calculate Cal1/Cal2};
  \node (maskSpectrum)     [DataReduction, below of=divideSpectra]   {Mask edges and atmospheric lines \\ and discard bad data};
  \node (lowPass)     [DataReduction, below of=maskSpectrum]   {Subtract low frequency signal};
  \node (sourceFinding)     [SourceFinding, below of=lowPass]   {Absorption line finding};    
  \draw[->]             (load) -- (extractSpectrum);
  \draw[->]      (extractSpectrum) -- (sumSpectrum);
  \draw[->]     (sumSpectrum) -- (divideSpectra);
  \draw[->]      (divideSpectra) -- (maskSpectrum);
  \draw[->]      (maskSpectrum) -- (lowPass);
  \draw[->]       (lowPass) -- (sourceFinding);
\end{tikzpicture}
\caption{A flowchart describing our methodology to efficiently process the large data volume of ALMACAL while maintaining the highest spectral resolution.}
\label{FigDataRedFlow}
\end{figure}

\begin{figure}
\includegraphics[width = \linewidth]{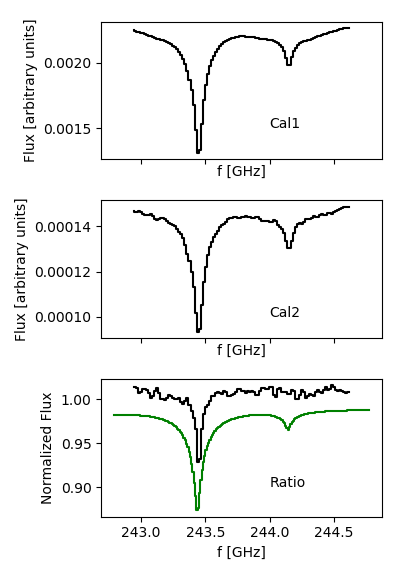}
\caption{Example of the first data reduction step to construct the ratios of two calibrator spectra observed in the same ALMA execution block. The top/middle panels show the spectra of calibrator 1 and 2, respectively, with arbitrary flux units. The bottom panel shows the ratio of the spectra of calibrator 1 and calibrator 2. The green line represents the atmospheric model as described in section \ref{SecDataRed}. This data processing reduces atmospheric line signatures.}
\label{FigRatio}
\end{figure}

\begin{figure}
\includegraphics[width=\linewidth]{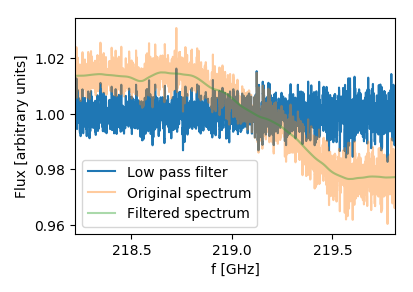}
\caption{Illustration of the application of a low-pass filter in the data processing. The orange spectrum shows the processed data before filtering, the green curve represents the same data after applying a low-pass filter maintaining features that are on scales larger than 200~km~s$^{-1}$. In blue, we show the resulting spectrum obtained by dividing the original spectrum by the low-pass curve. The resulting flat spectrum is then used as an input to the absorption line finder.}
\label{FigFilter}
\end{figure}

We extract from the ALMA archive all phase, amplitude and bandpass calibrator data from PI observations from Cycles 1 to 6, taken before the 4$^{\rm th}$ December 2018. We only consider data taken with the ALMA 12-m array. This amounts to observations of 880 calibrators. To determine the redshift of the calibrators, we use the compilation of redshifts from the database presented by \citet{Bonato2018}, combined with the updated redshift estimates of the Australian Telescope 20 GHz survey \citep[AT20G][]{Ekers2007} sources (\citet{Mahony2011} and E.~Mahony private comm.). For the remaining calibrators, we perform an additional query to the Simbad \citep{Wenger2000} and NED\footnote{ The NASA/IPAC Extragalactic Database (NED) is operated by the Jet Propulsion Laboratory, California Institute of Technology, under contract with the National Aeronautics and Space Administration.} databases. We note, however, that the accuracy of these redshifts might be limited. This results in redshift measurements of 622 calibrators. We test whether our samples of calibrators with and without redshift information are drawn from the same population of {\it WISE} colours (Band1 - Band4) and find that, based on a Kolmogorov- Smirnov test, we have to reject this hypothesis with a 96.4\% confidence level. However, we perform the line search on all quasar spectra irrespective of the availability of the redshift and therefore do not introduce a bias in our sample of calibrators. The line identification of absorption signals towards background quasars without redshift information is not straightforward since without a known redshift, the absorption line could both be intervening and associated to the background source. For such cases it would be important to obtain a redshift of the quasar using follow-up observations. Thus far, this is a hypothetical problem since we do not find any extragalactic absorption lines towards background quasars at unknown redshifts. In the following we discuss only the spectra of quasars with known redshifts.

We devise a new, optimised data processing strategy to handle the large data volume comprising several tens of thousands of spectra while maintaining the highest spectral resolution. This is necessary to keep the data volume manageable. For reference, we store in our ALMACAL archive fully-calibrated ms-files with a reduced spectral resolution which amounts to more than 26 Tb of data. The reduced spectral resolution of 15 kHz is too low to study absorption lines that are expected to be narrower than $40$~km~s$^{-1}$ \citep{Wiklind2018}. A schematic view of the data reduction chain is illustrated by Fig.~\ref{FigDataRedFlow}.
To this end, the spectrum of the calibrator is extracted directly from the $uv$ data, by fitting a point source model at the phase centre. For technical reasons, we extract the XX and YY polarisation data separately and add those in quadrature to obtain Stokes I spectra. We choose to only consider dual-polarization mode scans to keep the data retrieval simple and uniform, full polarisation data represents only a small fraction of the total ALMA archival data. For each calibrator observation, each spectral window is treated individually, resulting in a total of 28,644 spectra in our database. To remove unwanted structures from the spectra, we apply a bandpass correction by taking ratios of the spectra of pairs of calibrators from the same execution block. 
This procedure also removes some of the atmospheric absorption line signatures imprinted on the spectra. An example of this procedure is shown in Fig.~\ref{FigRatio}. For the vast majority of the calibrator observations, this simple algorithm results in flat spectra, apart from those narrow spectral regions that correspond to strong atmospheric absorption. If more than two calibrators were used in one observation, all possible combinations of calibrator spectra pairs are used to produce bandpass-calibrated spectra. 
This approach allows us to confirm a potential detection identified in one ratio-ed spectrum using a second ratio.

For further processing of the spectra, we mask 5 per cent of the channels on each end of the spectrum to remove edge effects. These edge channels are often strongly affected by non-flat bandpass effects. Furthermore, we mask a 0.2~GHz wide window centred on the central frequencies of the strongest H$_2$O, O$_2$ and O$_3$ atmospheric absorption lines identified from the ALMA atmosphere model provided by Juan Pardo\footnote{\url{https://almascience.nrao.edu/about-alma/atmosphere-model}}. Spectra covering more than one atmospheric transition are not further considered. Finally, contiguous parts of the spectra narrower than 15 per cent of the total spectrum bandwidth are discarded to ensure a possible detection of the full absorption line and the continuum. 

Despite the success of this simple algorithm, we observe that on several occasions, a second-order correction of the bandpass is required to remove all unwanted signal. To this end, we use a Butterworth low-pass filter developed as a maximally flat low-pass filter for signal processing. For each spectrum, we create a template of the spectrum including only structures wider than $200$~km~s$^{-1}$. The original spectrum is divided by this template of its low frequency shape. An example of this procedure is shown in Fig.~\ref{FigFilter}. 
We expect molecular absorption lines to be narrow ($<100$~km~s$^{-1}$) and moreover, the quality of the spectra does not allow us to search for wider spectral lines since they would be indistinguishable from instrumental artefacts and imperfect bandpass calibration.

\section{Analysis and Results}
\label{SecBlindDetec}

\begin{figure*}
\includegraphics[width = 0.49\linewidth]{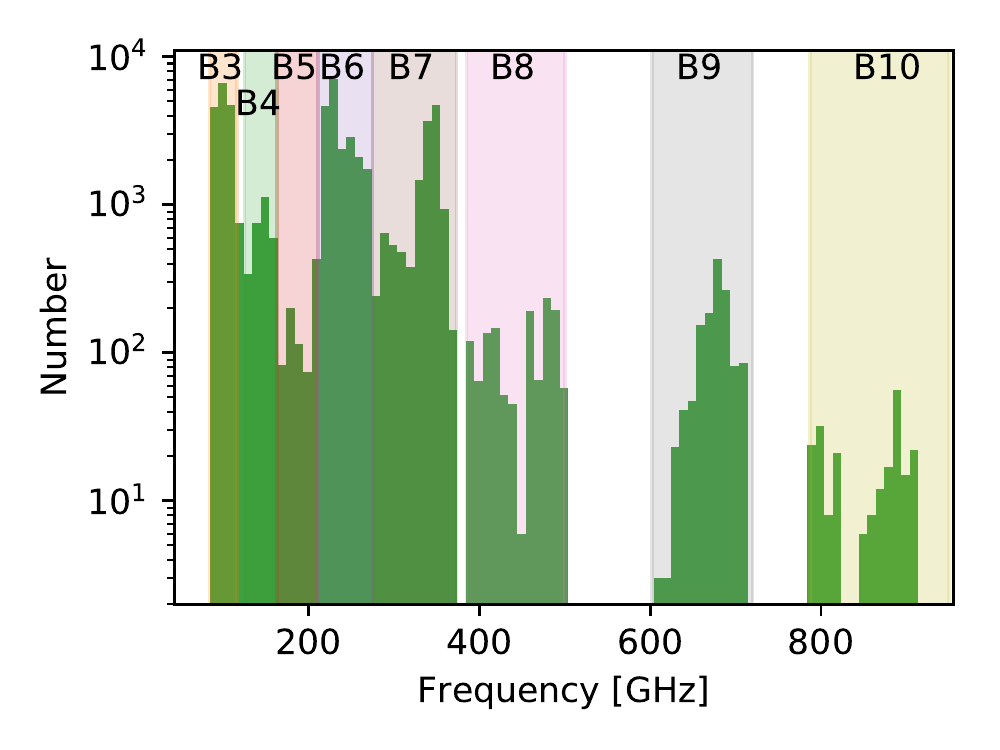}
\includegraphics[width = 0.49\linewidth]{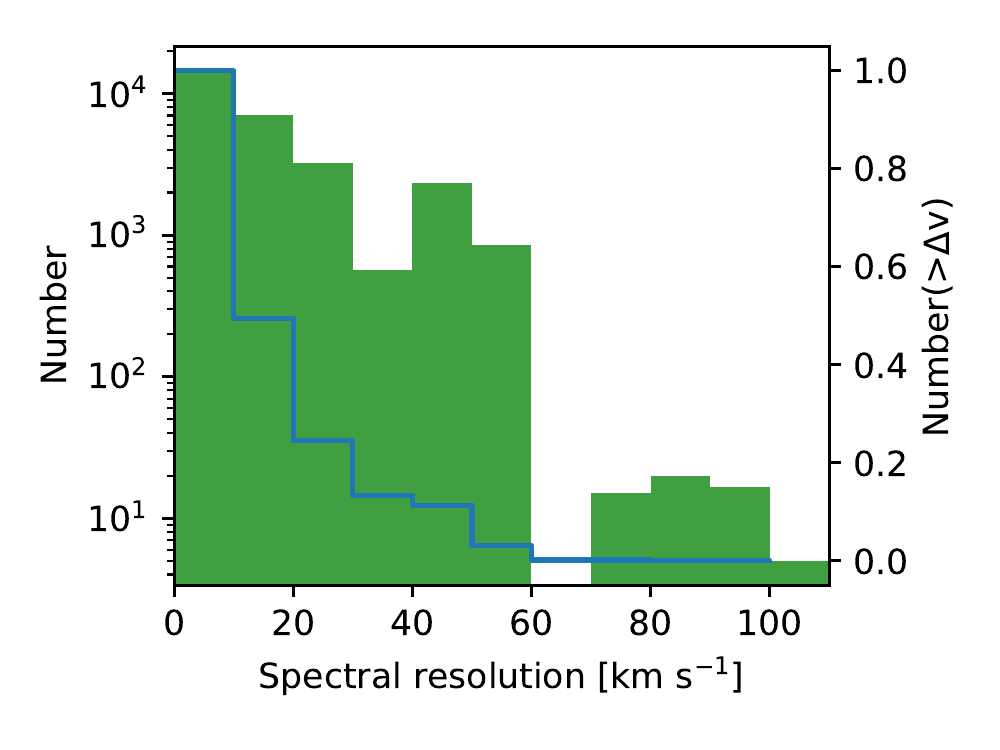}
\caption{Key quantities of the ALMACAL spectra. {\it Left:} Histogram of the observed frequencies with the ALMA observing bands overlaid. Most of the observations are taken in ALMA bands 3, 6, and 7. {\it Right:} Distribution of the velocity resolution of the ALMACAL spectra as a histogram and a cumulative plot. Half of the spectra have a resolution higher than 10 km s$^{-1}$. \label{FigDataDescr}}
\end{figure*}

\subsection{Blind Search for Intervening Absorbers}
\begin{figure*}
\includegraphics[width = 0.49\linewidth]{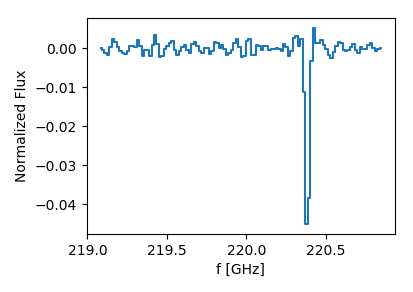}
\includegraphics[width = 0.49\linewidth]{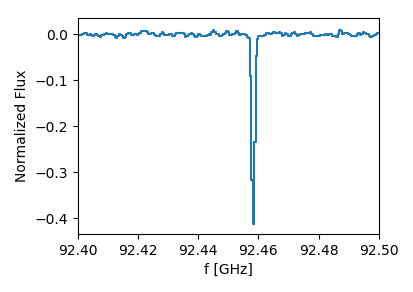}
\caption{Examples of detected absorption lines. Left: A Galactic absorption line detection in the spectrum of J1744-3116. The absorption line is $^{13}$CO(2-1) at 220.39 GHz that arises in the ISM of the Milky Way Galaxy. Right: Associated $^{12}$CO(1-0) absorption in the known molecular absorber J1415+1322 at z = 0.24671. While we find multiple Galactic absorption lines, no extragalactic intervening molecular absorbers are detected.\label{FigDetection}}
\end{figure*}

After the data processing described in the previous section and discarding spectra based on the criteria described above, we are left with 28,644 flattened spectra, cleaned from any unwanted atmospheric or instrumental effects, in which we search for absorption lines of astrophysical origin. The key properties of the data including observed frequencies and velocity resolution are shown in Fig.~\ref{FigDataDescr}. We note that the spectra have varying bandwidth and spectral resolution, ranging from 0.03~MHz to 2~GHz, and from 256 to 3840 channels, respectively. We devise a search algorithm based on a signal-to-noise threshold of $5\sigma$, and apply this algorithm to both the ratios and the inverted ratios to search for absorption in both calibrator spectra. Since the spectra show no significant bandpass variations based on manual inspection, we can apply a global $\sigma$-threshold for each individual spectrum. Before running the finder algorithm, we smooth each spectrum to increase the signal-to-noise ratio. We use a range of smoothing kernels between 10 and 190~\kms~in steps of 10~\kms. Furthermore, we only record detections if the signal is significant in two consecutive channels.

From this initial list of candidates, we remove all detections which occur within the lowest or the highest 5 per cent of channels to remove absorption line candidates for which we do not see the continuum on both sides. Furthermore, we discard detections that lie within 200~\kms\ of the velocities of the centres of known Galactic CO lines. We apply an additional manual cleaning of candidate detections that can be identified as obvious electronic artefacts (such as strong periodic signals). For the remaining candidates, we identify atmospheric and Galactic transitions by cross-matching the detected frequencies observed in multiple sight lines. Finally, we match the candidate list with a list of frequencies corresponding to rare molecular species from SPLATALOGUE \citep{Remijan2007} to filter out remaining absorption lines of Galactic origin. Examples of a detection of Galactic absorption and associated absorption are shown in Fig.~\ref{FigDetection}. The Galactic absorption lines will be the subject of a forthcoming paper. Additionally, we compare the redshift of the lowest possible CO transition with the calibrator redshift and exclude implausible lines (i.e. the redshift of the absorber would be higher than the redshift of the background quasar). After performing these checks, we are left with one significant detection of an extragalactic molecular absorption line shown in Fig.~\ref{FigDetection}, which we identify to be intrinsic CO(1-0) absorption in the spectrum of the background calibrator J1415+1320 ($z = 0.2467$) \citep{Wiklind1994}. This detection validates the robustness of our finding algorithm.

\subsection{The Column Density Distribution Function Based on Intervening Absorbers}
\label{SecAnalysis}

We calculate the column density distribution function from the sensitivity limits we reach in the calibrator spectra. To illustrate the potential of this method we derive predictions of the column density distribution function from the IllustrisTNG100 cosmological hydrodynamical simulation \citep{Pillepich2018, Naiman2018, Nelson2018b, Marinacci2018, Springel2018}.

\begin{figure*}
\includegraphics[width = 0.49\linewidth]{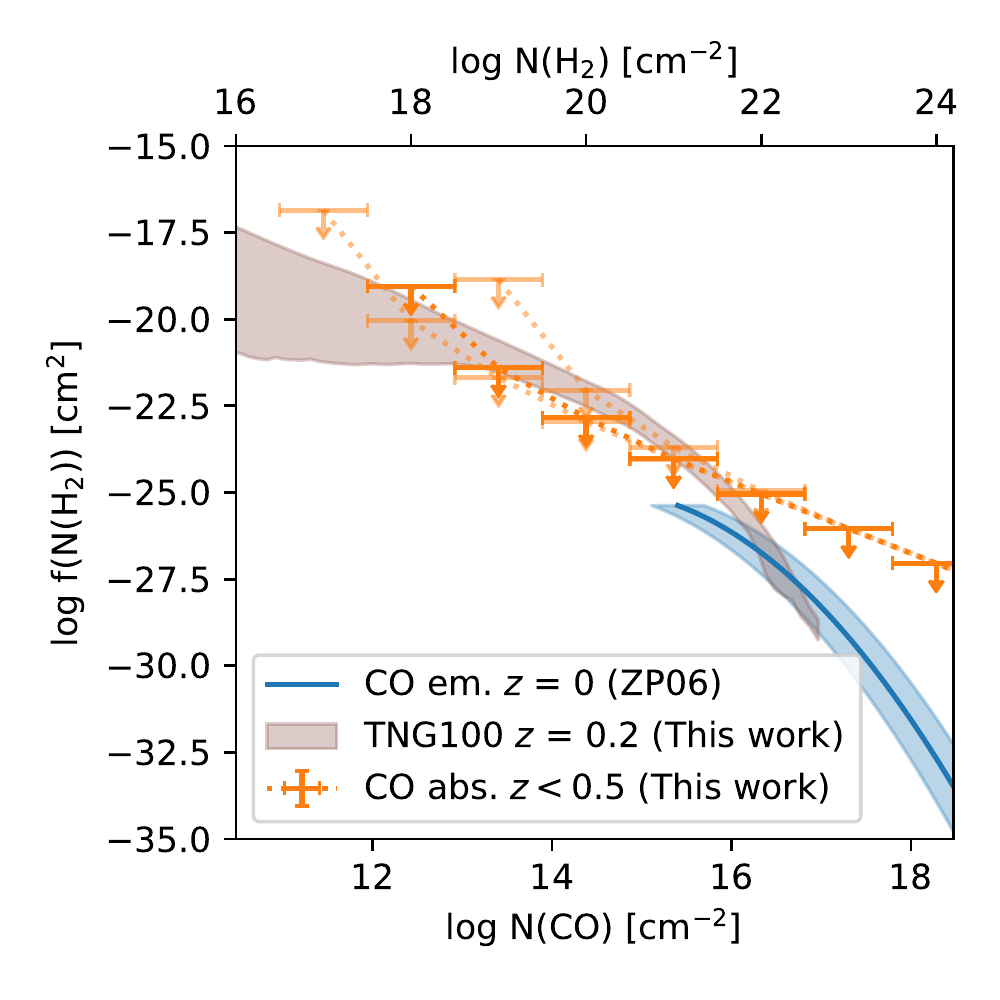}
\includegraphics[width = 0.49\linewidth]{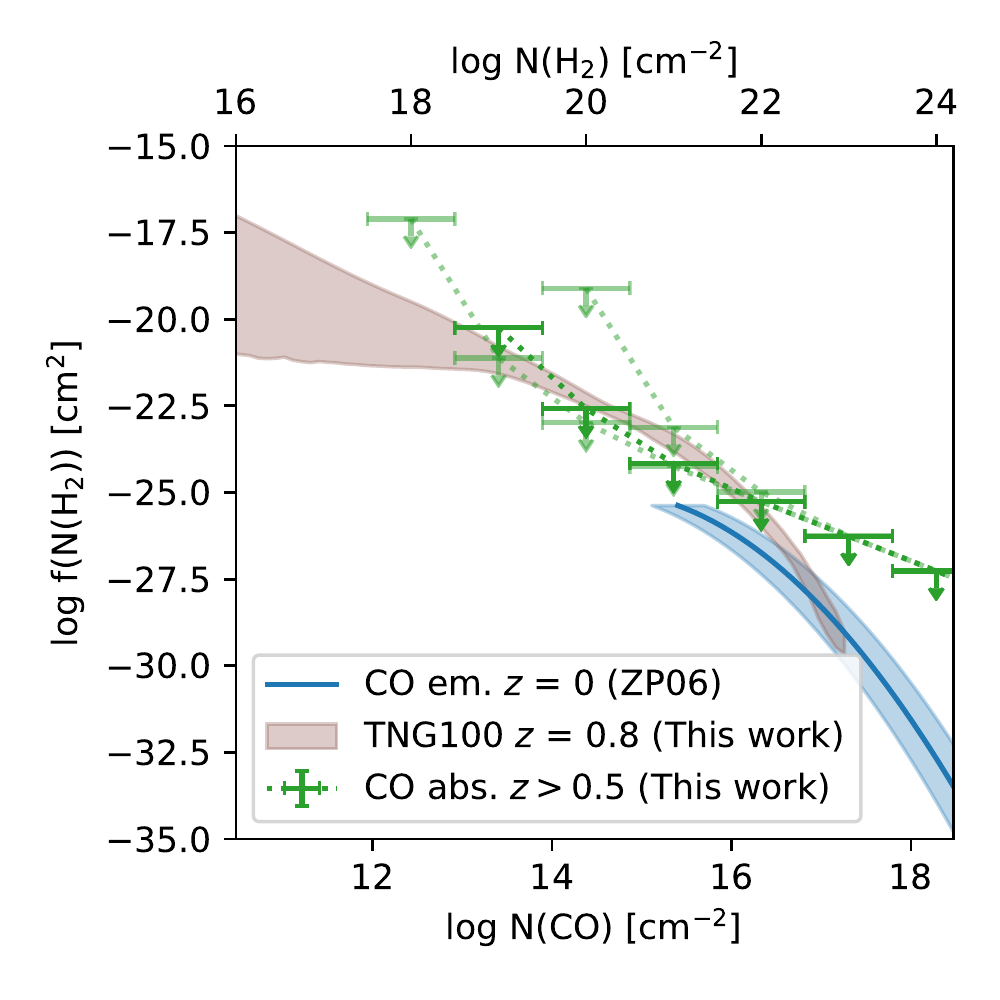}
\caption{CO column density distribution functions in the two redshift bins. The column densities are expressed in molecules cm$^{-2}$. The arrows indicate the upper limits from our ``blind'' CO absorber survey within $\Delta N = 1$~dex. The left panel corresponds to $z<0.5$ and the right panel to $z>0.5$. Light coloured limits reflect the uncertainty introduced by the CO-to-H$_2$ column density conversion factor. The blue line is the H$_2$ column density distribution function at $z=0$ based on CO emission line observations \citep{Zwaan2006}. The brown shaded region marks the predictions based on IllustrisTNG100 results with a variation of post processing recipes to illustrate the uncertainties (see Sec.~\ref{SecTNG100} for details). The top-axis shows the fiducial CO-to-H$_2$ conversion from \citet{Burgh2007}.}
\label{FigColDensDist}
\end{figure*}

\begin{table}
\caption{Redshift path surveyed, $\Delta z$, and comoving pathlength, $\Delta X$, for each CO transition in two distinct redshift ranges, $z < 0.5$ and $z > 0.5$. The cumulative redshift path surveyed, $\Delta z$, is 182.2 for CO transitions between $J = 1-0$ and $J = 5-4$.}
\label{TabLimColumnDensLow}
\centering
\begin{tabular}{l l l l l l l}
\hline \hline
CO & Redshift & $\Delta z$ & $\Delta X$ \\ 
transition&range & &\\
\hline
CO(1--0) & $<0.5$ & 48.4 & 61.2 \\
CO(2--1) & $<0.5$ & 13.4 & 16.3 \\
CO(3--2) & $<0.5$ & 20.4 & 28.6 \\
CO(4--3) & $<0.5$ & 9.8 & 14.6 \\
CO(5--4) & $<0.5$ & 1.4 & 2.1 \\
\hline
CO(1--0) & $>0.5$ & 0.0 & 0.0 \\
CO(2--1) & $>0.5$ & 34.1 & 80.8 \\
CO(3--2) & $>0.5$ & 18.5 & 42.0 \\
CO(4--3) & $>0.5$ & 18.8 & 41.5 \\
CO(5--4) & $>0.5$ & 17.5 & 40.0 \\
\hline \hline
\end{tabular}
\end{table}

\begin{figure*}
    \centering
    \includegraphics[width = 0.49\linewidth]{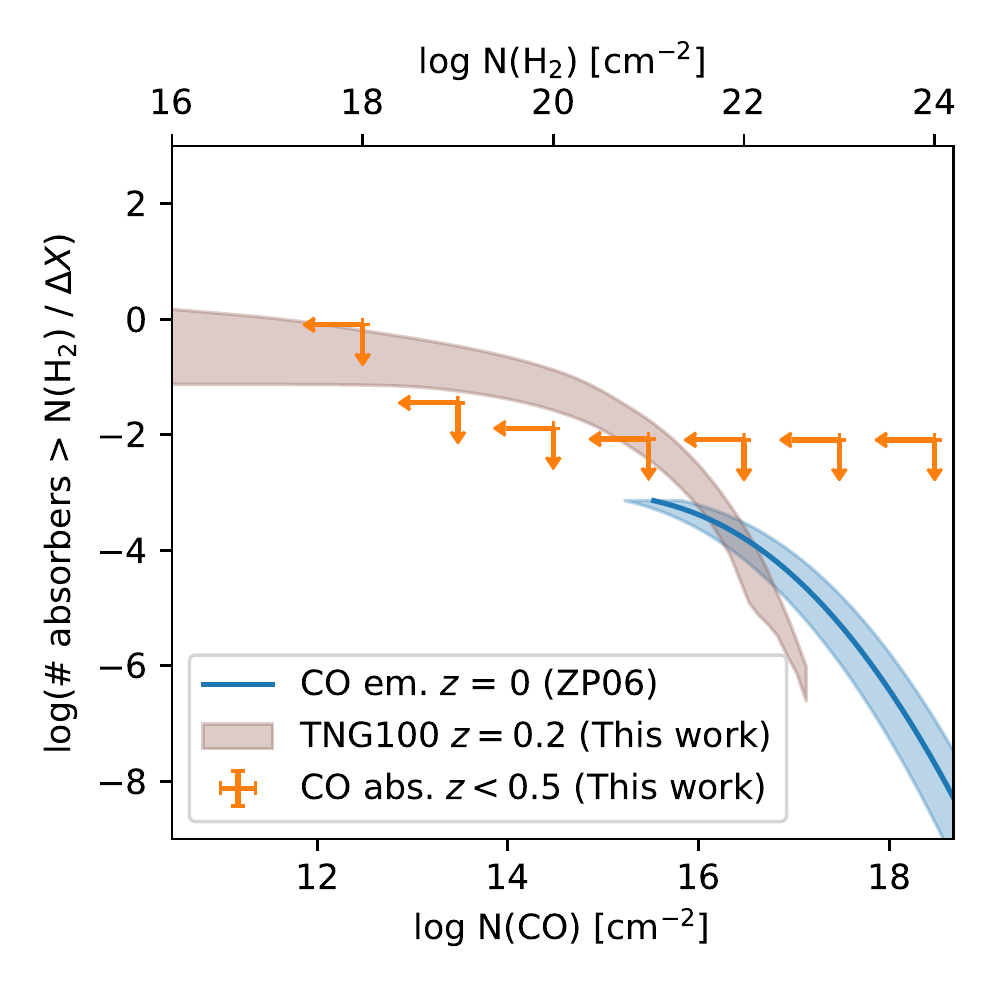}
    \includegraphics[width = 0.49\linewidth]{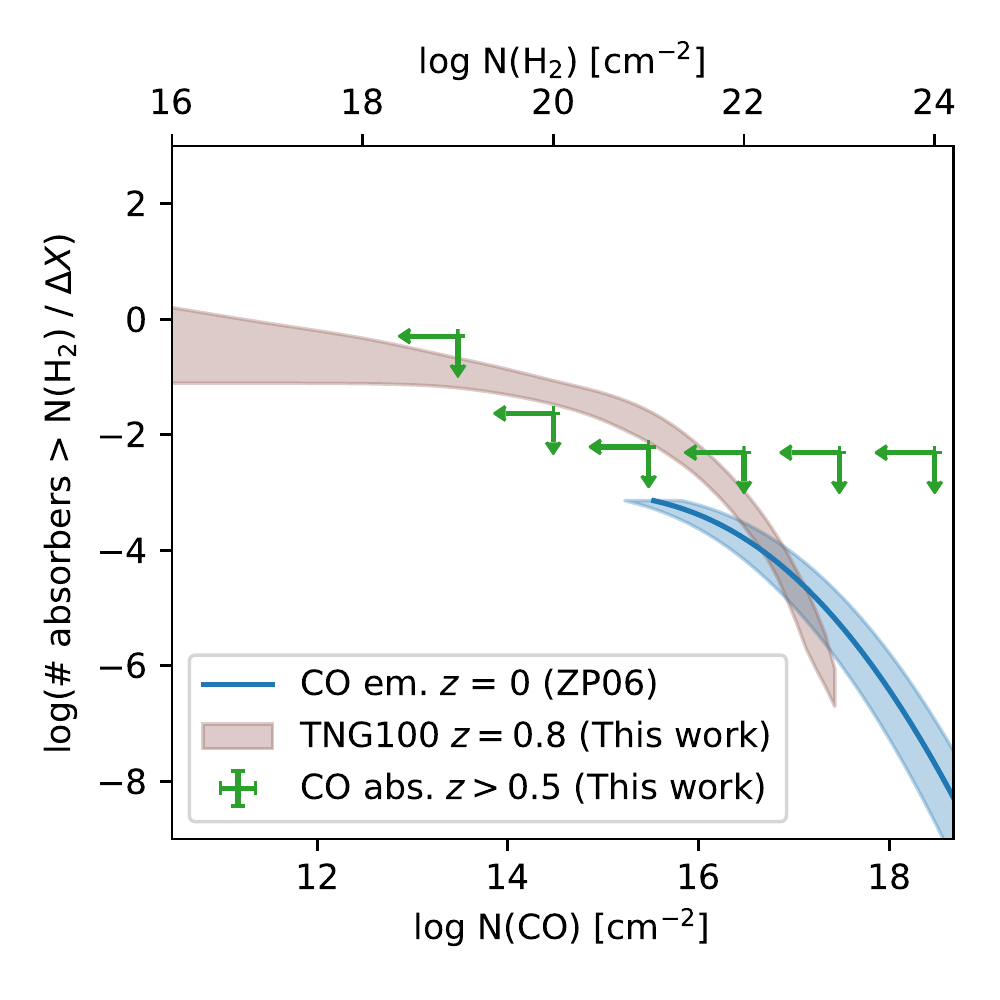}
    \caption{Cumulative number of molecular absorbers per comoving path length interval $\Delta X$ with a column density greater than $N$ (notations same as in Fig.~\ref{FigColDensDist}). This presentation of the column density distribution function is independent of the choice of a bin size, $\Delta N$. We use this bin-free representation to calculate limits on the molecular gas densities with redefined redshift bins by scaling of the functional form from \citet{Zwaan2006} to our upper limits (see Sec.~\ref{SecRhoH2} for details).}
    \label{FigCumNum}
\end{figure*}

We first calculate the redshift path probed by the 28,644 spectra included in our survey. For each spectrum, we compute the redshift path observed \citep[see e.g.][]{Zafar2013} given the observed frequencies and assuming CO transitions from $J = 1-0$ up to $J = 5-4$. The frequency coverage of the fully reduced and masked spectra is used for this calculation. The maximum probed redshift in each spectrum is set by the redshift of the calibrator. 
The cumulative redshift path surveyed, $\Delta z$, is 182.2 for CO transitions between $J = 1-0$ and $J = 5-4$. We further split the sample in two redshift ranges, at $z > 0.5$ and $0.5 < z < 1.7$ with mean redshifts of $z = 0.199$ and $z = 0.839$. The two subsamples are covering approximately the same pathlength of $\Delta z = 93.3$ at $z < 0.5$ and  $\Delta z = 88.9$ at $z>0.5$. Details of the redshift paths for each CO transition in the two sub-samples are listed in Table~\ref{TabLimColumnDensLow}.

We then calculate the limiting CO column densities probed in our survey following \citet{Mangum2015}. We assume an excitation temperature equal to the CMB temperature at the redshift probed with the spectrum, because this is the lowest possible temperature. The physical conditions of the molecular absorbing gas in the galaxy lensing the quasar PKS1830--211 were investigated by \citet{Muller2013}. They found that for polar molecules, the excitation temperature is close to that of the CMB at the corresponding redshift. A molecule like CO, on the other hand, is easier to excite due to its low electric dipole moment, and in general, we would not be able to constrain $T_{\rm ex}$ for CO to better than $T_{\rm cmb} < T_{\rm ex} < T_{\rm kin}$, without constraints from additional lines/species. Since we have no detections, we perform the calculation using the $5 \sigma$ level from each spectrum as the detection threshold and an expected FWHM of the absorption line of $40$~km~s$^{-1}$ \citep{Wiklind2018}. The CO column density limit is converted into a H$_2$ column density limit using a mean column density ratio of $N({\rm CO})/ N(\htwo) = 3 \times 10^{-6}$ \citep{Burgh2007}. In order to bound the large uncertainty on this conversion factor, we also present CO column densities derived with upper and lower limits of $10^{-5}$ to $10^{-7}$, respectively. The column density limits from non-detections are calculated for each observation using the corresponding frequency coverage and rms. We note that the column density ratio of $N({\rm CO})/ N(\htwo)$ over a large range of H$_2$ column densities is not constant \citep{Balashev2017}. However, with the currently available data this is challenging to quantify.

Next, we estimate the $5 \sigma$ limits on the column density distribution function following the definition \citep{Carswell1984}:

\begin{equation}
f(N(\htwo),X) dN(\htwo) \Delta X < \frac{1}{\Delta N(\htwo) \Delta X} dN(\htwo) \Delta X,
\end{equation}

where the number of absorbers detected within the column density range $\Delta N(\text{H}_2)$ is less than one. $\Delta X$ is the comoving pathlength for the specific column density under consideration. The comoving pathlength ensures that for a constant physical size and comoving number density, the absorbers have a constant $f(N(\text{H}_2), X) $ \mbox{\citep{Bahcall1969}}. The comoving pathlength of a single sightline, $\Delta X_i$, is defined as follows:

\begin{equation}
\Delta X = \frac{H_0}{H(z)} (1 + z)^2 \text{d} z
\end{equation}

\begin{equation}
\Delta X_i = \int_{z_{\rm min}}^{z_{\rm max}} \text{d} X = \int_{z_{\rm min}}^{z_{\rm max}} \frac{(1 + z)^2}{\sqrt{ \Omega_{\Lambda} + \Omega_{\rm M} \times (1 + z)^3}} \text{d} z.
\end{equation}

The limiting column densities and covered path length are then combined for the whole survey. 

The non-detections from our survey translate to upper limits on the column density distribution function. However, in the definition of $f(N(\text{H}_2), X)$ the choice of the bin size influences the values of $f(N(\text{H}_2), X)$ upper limits in the case of non-detections. Here, we use a bin width of $\Delta N = 1$~dex, as it is common practice in \hi\ absorption line studies \citep[e.g.][]{Peroux2003}. The resulting upper limits on the column density distribution function are shown in Fig.~\ref{FigColDensDist} for the two redshift ranges. 

To remedy the dependence on the bin size, we also calculate the cumulative number of absorbers per $\Delta X$ \mbox{\citep{Peroux2003}} as a function of column density, which is independent of the binning choice (see Fig.~\ref{FigCumNum}). We also calculate the cumulative number of absorbers expected based on the results from BIMA SONG observations of local star-forming galaxies \mbox{\citep{Zwaan2006}} for comparison.  

\subsection{Predicting the Column Density Distribution Function from IllustrisTNG}
\label{SecTNG100}

\begin{figure}
    \centering
    \includegraphics[width = 0.8\linewidth]{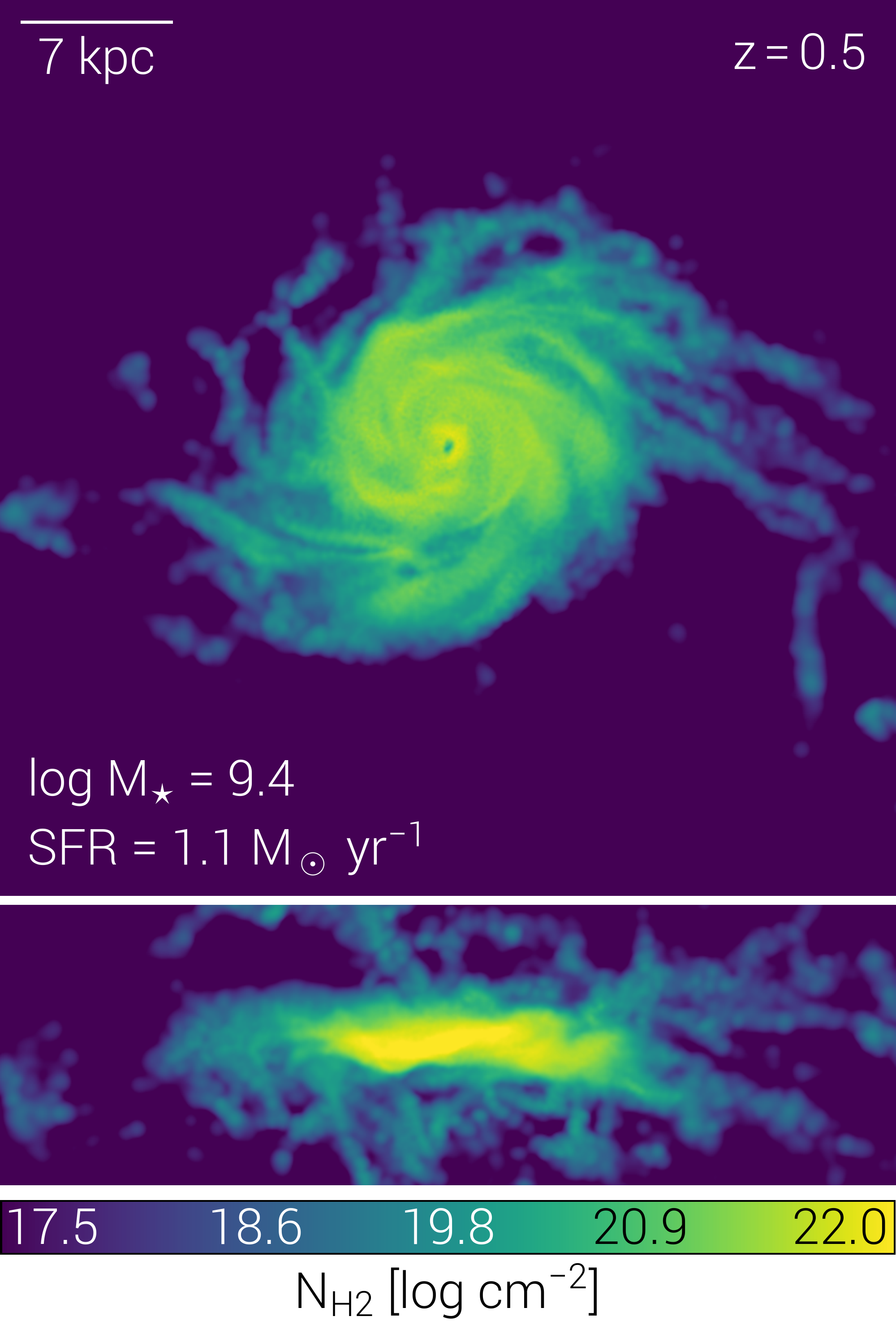}
    \caption{An example of a molecular gas disk in a z=0.5 galaxy (top panel: face-on, bottom panel: edge-on view) with \mbox{$M_{\star} = 10^{9.4} {\rm M}_{\sun}$} and \mbox{${\rm SFR} = 1.1 {\rm M}_{\sun} {\rm yr}^{-1}$} from post processing of the IllustrisTNG100 simulation. The highest column densities are only observed in edge-on disks, while intermediate column densities are predicted out to radii of $\sim$10~kpc in other viewing directions.}
    \label{FigTNGmolDisk}
\end{figure}

From a modelling point of view, the molecular phase of the cold gas is challenging to assess because of the complexity of the physics involved and because it requires sub-grid modelling to capture the unresolved physics. Semi-analytical techniques of pressure-based models \citep{Blitz2006, Gnedin2011, Krumholz2013} are used to split the cold hydrogen from hydrodynamical simulations (such as the EAGLE or IllustrisTNG) into its atomic and molecular components \citep{Obreschkow2009, Popping2014, Lagos2015, Chen2018a}. 

\begin{figure}
    \centering
    \includegraphics[width = \linewidth]{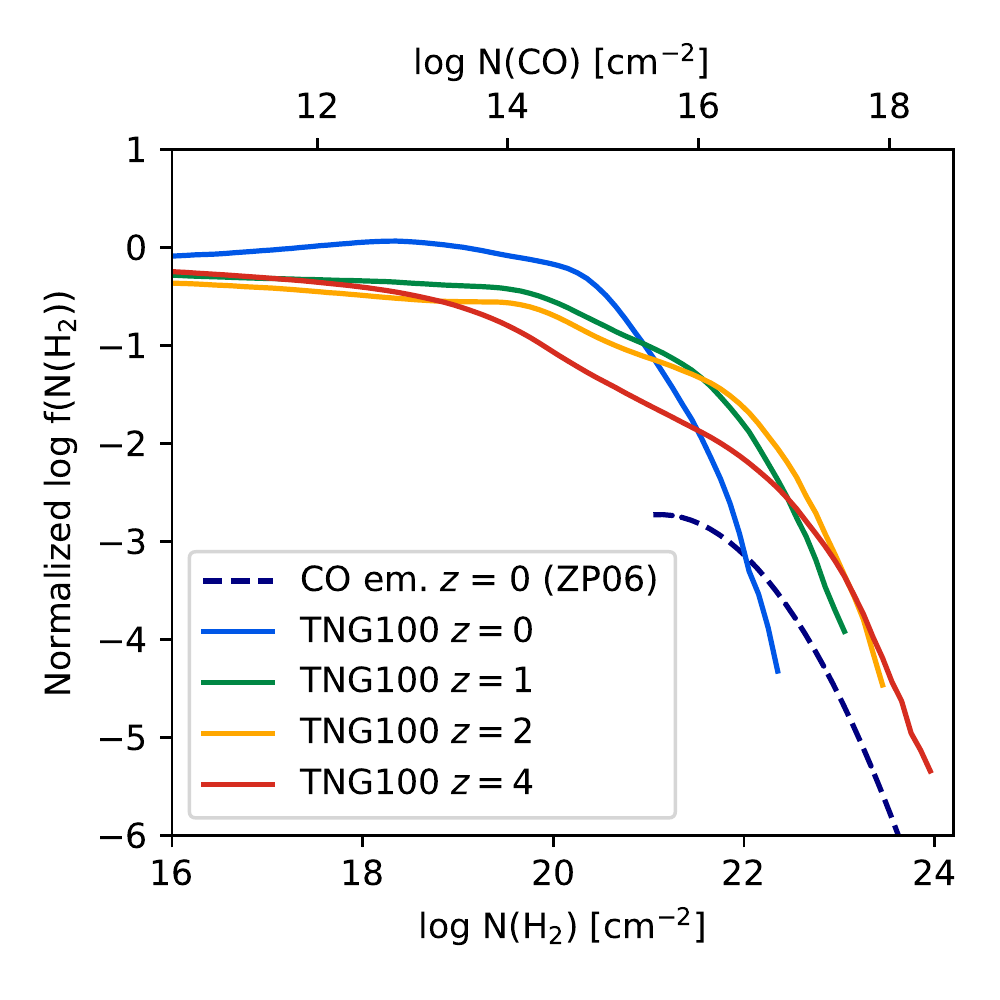}
    \caption{Redshift evolution of the H$_2$ column density distribution function as predicted from the IllustrisTNG simulation (solid lines) and from observations at $z = 0$ by \citet{Zwaan2006} (dashed line) normalized by a power law function fitting the low column density end of the predictions. We find that the column density distribution functions determined from the post processing of IllustrisTNG results at different redshifts predict an increasing number of high column density absorbers at high redshift.}
    \label{FigCDDFzevoTNG}
\end{figure}

Here we use the TNG100 volume of the IllustrisTNG simulations \citep{Pillepich2018, Naiman2018, Nelson2018b, Marinacci2018, Springel2018} through its publicly available data \citep{Nelson2018c} in order to compare our observations against the theoretical expectation for the $\htwo$ column density distribution function. An example of the column density map in a typical galaxy from the simulations is shown in Fig.~\ref{FigTNGmolDisk}. We construct the column density distribution function at the mean redshift of the two subsamples ($z=0.199, 0.839$) using the $\htwo$ modeling methodology of \citet{Popping2019} (see also \citet{Stevens2019, Diemer2019} for assessments of the \hi~and $\htwo$ outcomes of TNG) and the column density distribution function gridding procedure as described in \citet{Nelson2019}. The column density is integrated over a path length of 10 cMpc $h^{-1}$. In order to assess the sensitivity of our result to various physical and numerical choices, we present a band which encompasses several different column density distribution function calculations, which vary the $\htwo$ model employed (three versions), the projection depth / effective path length (five values), different grid sizes for the computation of the column density (three values), and assumptions on the $\htwo$ contents of star-forming versus non-star-forming gas cells. In Fig.~\ref{FigCDDFzevoTNG}, we show the expected evolution of the column density distribution function with redshift. We find that an increasing number of high column density absorbers at high redshift is expected. On the low column density end, on the other hand, the number of absorbers is almost constant over $z = 4 - 1$ and increases at $z = 1$. We also note that in the IllustrisTNG100 results, high column densities observed by \citet{Zwaan2006} are not reproduced. This may be due in part to the limited volume in the simulation. Furthermore, the simulations predict more low column density molecular gas compared to the observations. The expected error range shown in Fig.~\ref{FigColDensDist} also applies to this plot, but is not shown for reasons of clarity. This will not explain the full discrepancy. Additional uncertainties affecting the comparison between the simulations and observations are the assigning of a molecular gas fraction to the gas cells in the post processing as well as the CO-to-H$_2$ conversion factor used in the observations. Furthermore, \citet{Diemer2019} compare H$_2$ half mass radii in IllustrisTNG, relative to stellar half-mass radii, to the EDGE-CALIFA survey (based on CO), finding that although both TNG and EDGE-CALIFA have a majority of galaxies with $R_{\rm half,H2}/R_{\rm half,*} \sim 1$, the median ratio in TNG is approximately 30\% larger, though with a large dependence on the invoked HI/H$_2$ model. However, this degree of difference in H$_2$ extents cannot fully explain the large H$_2$ column differences seen here.

The prediction of the column density distribution function and the cumulative number of absorbers are shown in Fig.~\ref{FigColDensDist} and \ref{FigCumNum}. We have conducted the same analysis on the results from the EAGLE simulation \citep{Schaye2015, Crain2015} and find that the qualitative expectations for the column density distribution function are in line with those from IllustrisTNG.

\subsection{Cosmic Evolution of the Molecular Gas Mass Density}
\label{SecRhoH2}

\begin{figure*}
    \centering
    \includegraphics[width = 0.8\linewidth]{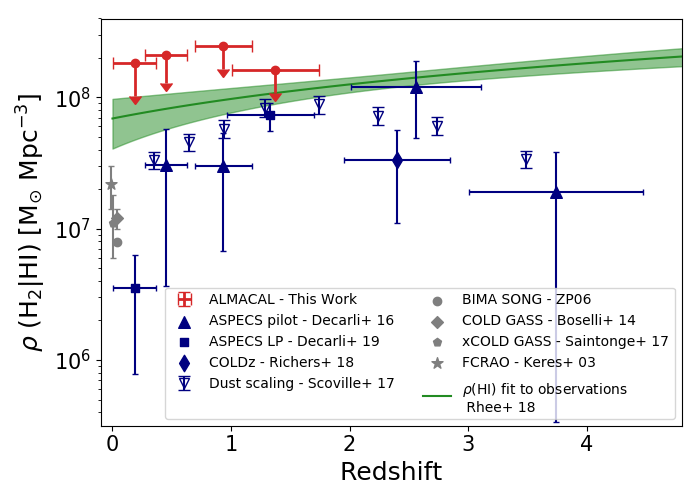}
    \caption{Cosmic evolution of the molecular and atomic gas densities. For $\rho ( {\rm H}_2)$, our limits are consistent with the measurements at $z = 0$ of \citet{Keres2003, Zwaan2006, Boselli2014, Saintonge2017} and the results of \citet{Decarli2016, Decarli2019, Scoville2017}. The sensitivity to low column density of the absorption line technique combined with the ALMACAL survey being unaffected by cosmic variance emphasize power of this complementary study to probe the cosmic evolution of the molecular gas mass density. A fit to $\rho ({\rm HI})$ observations is shown as a solid line \citep{Rhee2018}.}
    \label{fig:rhoComp}
\end{figure*}

Finally, we calculate the molecular gas mass density $\rho ( {\rm H}_2)$ from the cumulative number of absorbers per $\Delta X$. We use the functional form of the cumulative number of absorbers per $\Delta X$ from \citet{Zwaan2006} at $z \sim 0$ as a proxy. We scale it to our upper limits and integrate over differential number of absorbers multiplied by the respective column density. \citet{Zwaan2006} found that the contribution of low column density absorbers with \mbox{$\log(N({\rm H}_2)) < 21$} to the total molecular gas mass at $z \sim 0$ is only 3 per cent. Therefore, we integrate only  column densities log($N({\rm \htwo})$) $>$ 21. Since we aim at a comparison with other surveys, we define similar redshift bins to those introduced by the ASPECS survey \citep{Decarli2016, Decarli2019} for this calculation.
The resulting limits are shown in Fig.~\ref{fig:rhoComp}, where we also report measurements from the literature based on the same cosmology and CO-to-H$_2$ conversion factor. 

\section{Discussion}
\label{SecDiscussion}

With our ``blind'' survey for intervening molecular absorbers we put significantly improved constraints on the column density distribution function of molecular gas beyond $z\sim 0$. We compare our upper limits with the measurements at $z \sim 0$ presented by \citet{Zwaan2006} in Fig.~\ref{FigColDensDist} and find that our limits at $0<z<0.5$ and $0.5 < z < 1.7$ are consistent with the column density distribution function measurement at $z = 0$. The depth of our data translates to five orders of magnitude lower column densities than probed by \citet{Zwaan2006}. In addition, the absorption technique with a sensitivity independent of redshift in principle allows us to measure the redshift evolution of the column density distribution function. 

We calculate limits on the limiting cross section of the molecular gas per galaxy based on the non-detection and the surveyed redshift path. We assume a \citet{Schechter1976} galaxy luminosity function and a uniform and spherically symmetric distribution of molecular gas and follow the description in \citet{Peroux2005}. 
We derive a maximum radius of 4.8~kpc at $z < 0.5$ and 4.6~kpc at $z > 0.5$. \citet{Zwaan2006} find a median impact parameter of $N({\rm H_2}) > 10^{21} \;{\rm cm}^{-2}$ of 2.5~kpc, consistent with our upper limits. Our results provide statistical evidence that molecular gas around galaxies have a limited extent, well below the typical size of CGM regions. It however does not exclude that the CGM may contain more clumpy molecular gas. 

Compared to the predictions from IllustrisTNG from Sec.~\ref{SecTNG100} presented in Fig.~\ref{FigColDensDist} our limits are already close, within ~1 dex, of the expected value of $f(N(\text{H}_2), X)$ at low column densities. 
 The sensitivity reached in our survey is comparable to the column densities predicted by the simulations.  
However, uncertainties in this comparison are still large in both observations and simulations. The observations on the one hand involve a conversion from measured CO column densities to $\htwo$ column densities. Cosmological simulations on the other hand are lacking the resolution and associated small-scale physics to follow molecular cloud formation and rely on sub-grid physics models. Furthermore, the uncertainty of the molecular gas fraction in gas cells assigned in the post processing and the too extended H$_2$ disk sizes as measured by their half mass radii in IllustrisTNG \citep{Diemer2019} make a direct comparison challenging. Improvements on both sides are necessary to further explore the molecular column density distribution. 

Fig.~\ref{fig:rhoComp} shows the cosmic evolution of the cold gas in the Universe. Dedicated efforts to measure the cosmic evolution of the molecular gas mass density from deep CO emission line observations by the ASPECS and COLDz surveys have provided the first measurements of $\rho ( {\rm H}_2)$ over a large redshift range \citep{Decarli2016, Decarli2019, Riechers2018}. Uncertainties in the $\rho ( {\rm H}_2)$ measurements, such as those related to uncertain CO excitation, completeness errors, and redshift errors are discussed by the authors of these studies. 
At least of similar importance for deep surveys with small fields of view such as ASPECS and COLDz are the effects of cosmic variance on $\rho (\htwo)$ measurements. This has been shown to be particularly important at low redshift \citep{Popping2019}. A complementary approach using the dust emission yields comparable results on  $\rho (\htwo)$ \citep{Scoville2017}. In our absorption-based study presented this paper, we provide new upper limits free from cosmic variance effects. Our limits are consistent with the measurements at z = 0 of \citet{Keres2003, Zwaan2006, Boselli2014, Saintonge2017} and supportive of the results of \citet{Decarli2016, Decarli2019, Scoville2017}. 
A fit to $\rho ({\rm HI})$ observations is shown based on \citet{Rhee2018}. These results show that the amount of cold gas in its atomic form is only a few times higher than that in its molecular phase from $z \sim 3$ to $z \sim 0$, implying that the decrease of $\htwo$ density is faster than for \hi~towards late times. While the SFH evolves by a factor 20--30 from $z=2$ to present day, $\rho ( {\rm H}_2)$ decreases by one order of magnitude in the same time lapse and $\rho ({\rm HI})$ by less than 15 per cent. These findings indicate that H$_2$ is being consumed faster than \hi\ can replenish it unless it is constantly fed. The dramatic decrease of the cosmic star-formation rate density might therefore arise from a shortfall of molecular gas supply. On the contrary the MUFASA simulation predicts a shallower evolution of the molecular gas mass density than indicated by the observations \citep{Dave2017}.

Future blind absorption line surveys will offer more stringent constraints on the evolution of the cosmic molecular gas mass density by either moving to higher redshifts, where more high column density absorbers are predicted per $dz$, or by increasing the surveyed redshift path. For our current survey we would be sensitive to the measurements from ASPECS if we would cover a 1.3 times larger comoving redshift path ($\Delta X \sim 290$). To put this into perspective, it is important to realize that the ALMACAL results presented in this paper are based on more observing time than the sum of all ALMA Large Programs from Cycles 4 to 7. ALMACAL is an ongoing survey, so more redshift pathlength is accumulated continuously. But even a modest increase of a factor of two will take several years of observing. Another significant improvement in the covered redshift pathlength would be achieved by increasing the instantaneous frequency coverage of ALMA observations from its current 8~GHz per polarisation to at least 16~GHz, as is recommended in the ALMA development roadmap \citep{Carpenter2019}. Apart from this technological improvement, an increase of the redshift path could be achieved by measuring more optical redshifts for ALMA calibrator sources, which is under way. However, the uncertainties introduced by lensing of the background quasar by the foreground absorber will remain a systematic issue.

\section{Summary and Conclusions}
\label{SecConclusions}

\begin{table}
\caption{Derived upper limits on the cosmic molecular gas mass density. \label{TabRhoH2Limits}}
\centering
\begin{tabular}{cl}
\hline
$z$ & $\rho({\rm H}_2) [\text{M}_{\sun} \text{Mpc}^{-3}]$\\
\hline
0.003 -- 0.369 & $< 10^{8.26}$ \\
0.2713 -- 0.6306 & $< 10^{8.32}$\\
0.6950 -- 1.1744 & $< 10^{8.39}$\\
1.006 -- 1.738 & $< 10^{8.21}$\\
\hline
\end{tabular}
\end{table}

We present constraints on the cosmic evolution of the molecular gas density of the Universe from a ``blind'' search for extragalactic intervening molecular absorbers using the ALMACAL survey. The novelty of the approach resides in {\em i)} its redshift-independent sensitivity, {\em ii)} its ability to reach low gas densities, and {\em iii)} the fact that it overcomes cosmic variance effects. Our survey is sensitive to column densities as low as N(CO)>10$^{11}$ cm$^{-2}$ (${\rm N}(\htwo) > 10^{16} {\rm cm}^{-2}$). This is five orders of magnitude lower than probed in previous surveys \citep{Zwaan2006, Kanekar2014}. 

To keep the data reduction simple and uniform, we use an simple data processing method to handle the large data volume while maintaining the data at its highest spectral resolution. The resulting sample of 622 unique quasar spectra is searched ``blindly" for CO absorption lines. 
At $z<0.5$, we survey a total pathlength of $\Delta z = 93$ and a total comoving pathlength of $\Delta X = 123$. At $z>0.5$, $\Delta z = 89$ and $\Delta X = 205$. The large path length surveyed allows us to put constraints on the CO column density distribution functions at $z<0.5$ and $z>0.5$. While we detect multiple Galactic absorption lines and one known extragalactic intrinsic absorber, no extragalactic intervening molecular absorbers have been found. 

The upper limits on the molecular mass density reported in this survey are presented in Table~\ref{TabRhoH2Limits}. 
These upper limits are consistent with previous surveys. Together, these findings indicate that the dramatic decrease of the star-formation rate history might arise from a shortfall of molecular gas supply. Our limits add a constraint on the contribution from low column density molecular hydrogen. In addition, the new absorption line technique offers a characterization of cosmic variance issues possibly affecting emission surveys \citep{Popping2019}. 
    
We present the theoretical estimates of the molecular gas column density distribution from post-processing of the IllustrisTNG results. These estimates are consistent with our observational upper limits. 
However, both are subject to systematic uncertainties. Both a better understanding of the CO-to-H$_2$ conversion factor and advances in the modelling of molecular gas in cosmological simulations will decrease the uncertainties.
    
To put stronger constraints on the evolution of the molecular gas mass, a significant improvement on the redshift path covered per observation with ALMA is needed. This will occur naturally over time and will be accelerated by the proposed technological upgrades. Further improvement will result from the measurement of background quasar redshifts.

\section*{Acknowledgements}

The authors thank Ryan J. Cooke for useful discussions, Jonghwan Rhee for providing the fit to $\rho ({\rm H {\sc I}})$ and Sergei Balashev for discussions on the CO-to-H$_2$ column density conversion.
AK acknowledges support from the STFC grant ST/P000541/1 and Durham University. CP thanks the Alexander von Humboldt Foundation for the granting of a Bessel Research Award held at MPA. IRS acknowledges support from the ERC Advanced Grant DUSTYGAL (321334), a Royal Society/Wolfson Merit Award and STFC (ST/P000541/1). RD thanks the Alexander von Humboldt Foundation for support. ALMA is a partnership of ESO (representing its member states), NSF (USA), and NINS (Japan), together with NRC (Canada), NSC and ASIAA (Taiwan), and KASI (Republic of Korea), in cooperation with the Republic of Chile. The Joint ALMA Observatory is operated by ESO, AUI/NRAO, and NAOJ. This research has made use of the SIMBAD database, operated at CDS, Strasbourg, France . This research has made use of the NASA/IPAC Extragalactic Database (NED), which is operated by the Jet Propulsion Laboratory, California Institute of Technology, under contract with the National Aeronautics and Space Administration.




\bibliographystyle{mnras}
\bibliography{MyLibraryNoLink} 




%
%


\bsp	
\label{lastpage}
\end{document}